\documentclass[fleqn,10pt]{SelfArx} 

\usepackage[english]{babel} 
\usepackage{fancyhdr}
\usepackage{multirow}

\definecolor{color1}{RGB}{0,0,90} 
\definecolor{color2}{RGB}{0,20,20} 

\usepackage{hyperref} 

\hypersetup{
	hidelinks,
	colorlinks,
	breaklinks=true,
	urlcolor=color2,
	citecolor=color1,
	linkcolor=color1,
	bookmarksopen=false,
	pdftitle={Title},
	pdfauthor={Author},
}

\JournalInfo{February, 2021} 
\Archive{Research Article} 

\PaperTitle{Modeling and prediction of COVID-19 in the United States considering population behavior and vaccination} 

\Authors{Thomas Usherwood\textsuperscript{1,2,+}, Zachary LaJoie\textsuperscript{1,2,+},Vikas Srivastava\textsuperscript{1,2}*} 
\affiliation{\textsuperscript{1}\textit{School of Engineering, Brown University, Providence, RI 02912, USA}} 
\affiliation{\textsuperscript{2}\textit{Center for Biomedical Engineering, Brown University, Providence, RI 02912, USA}} 
\affiliation{\textsuperscript{+}\textit{Both authors contributed equally}} 
\affiliation{ *\textbf{Corresponding author:} Vikas Srivastava\\
Email: vikas\_srivastava@brown.edu\\Brown University, 184 Hope Street, Box D, Providence, RI 02912} 

\newcommand{\keywordname}{Keywords} 

\Abstract{COVID-19 has devastated the entire global community. Vaccines present an opportunity to mitigate the pandemic; however, the effect of vaccination coupled with the behavioral response of the population is not well understood. We propose a model that incorporates two important dynamically varying population behaviors: level of caution and sense of safety. Level of caution increases with the number of infectious cases, while an increasing sense of safety with increased vaccination lowers precautionary behaviors. To the best of our knowledge, this is the first modeling approach that can effectively reproduce the complete time history of COVID-19 infections for various regions of the United States and provides relatable measures of dynamic changes in the population behavior and disease transmission rates. We propose a  parameter $d_I$ as a direct measure of a population's caution against an infectious disease, that can be obtained from the ongoing new infectious cases. The model provides a method for quantitative measure of critical infectious disease attributes for a population including highest disease transmission rate, effective disease transmission rate, and disease related precautionary behavior. We predict future COVID-19 pandemic trends in the United States accounting for vaccine rollout and behavioral response. Although a high rate of vaccination is critical to quickly end the pandemic, we find that a return towards pre-pandemic social behavior due to increased sense of safety during vaccine deployment, can cause an alarming surge in infections. Our results indicate that at the current rate of vaccination, the new infection cases for COVID-19 in the United States will approach zero by the end of August 2021. The model can be used for predicting future epidemic and pandemic dynamics before and during vaccination. 
}

\begin{document}

\maketitle 


\thispagestyle{empty} 


\section*{Introduction} 

\addcontentsline{toc}{section}{Introduction} 

Coronavirus Disease 2019 (COVID-19) began as a localized outbreak in Wuhan, China in December 2019 and quickly spread internationally to become a global pandemic. More than a year later, over 113 million people have become infected with COVID-19 with more than 2.5 million deaths worldwide \cite{WHO_dashboard}. To combat the spread of this virus, the Pfizer – BioNTech COVID-19 vaccine was approved in the United Kingdom on December 2, 2020 \cite{ledford_uk_2020} , and the Pfizer – BioNTech and Moderna vaccines were subsequently approved for emergency use authorization in the United States \cite{commissioner_covid-19_2021}. In light of these recent developments, the potential impact of the distribution for COVID-19 vaccines is tremendous. Giving the public, health officials and government, model-based trend predictions and additional guidance on effective vaccine distribution and potential problems is critical. 

In the last year, many papers have been published on modeling the COVID-19 pandemic \cite{Bertozzi2020, Estrada2020, Giordano2020, Kaxiras2020, Korolev2021, Tsay2020, Kucharski2020, He2020, Prem2020, Kennedy2020}. Several modeling studies are based on differential equation compartment models involving compartments for susceptible, infectious, and recovered individuals, commonly referred to as SIR models \cite{roda2020, Liu2020, Daron2020, Postnikov2020, Brauer2017, KermackMcKendrick1927}. Introducing additional compartments allows researchers to study the effect of vaccination and examine how to optimally distribute a vaccine. Matrajt et al. \cite{matrajt_vaccine_2020} used an age-stratified population to determine the consequence of vaccine effectiveness and population coverage of the vaccine to indicate the optimal vaccine allocation. Bubar et al. \cite{bubar_model-informed_2020} accounted for the possibility of ruling out individuals with antibodies from receiving the vaccine using a serological test, and added an age-dependent effectiveness of the vaccine. Effects of vaccination have been examined in past outbreaks such as the 2009 H1N1 Swine Flu outbreak and the 2014-2016 Ebola epidemic \cite{Feng2011, Scherer2002, chowell_vaccination_2019, lee_modelling_2013, larson_modeling_2012, potluri_impact_2020}.  These studies have introduced population compartments that separate the population by their location and give insights into what locations should receive vaccines first \cite{yu_efficient_2016}.  

A critical aspect of COVID-19 vaccination that remains unexplored is a population's behavioral changes during the prolonged period of vaccination. 
While behavioral responses have not been addressed with respect to vaccines, efforts have been made to study the effects of non vaccine related behavioral changes for previous pandemics. These studies vary from the models on the effectiveness of social measures like quarantining and social distancing \cite{perra_towards_2011, sardar_assessment_2020, kim_prediction_2020} to characterizing the nature of spread of the disease \cite{sarkar_modeling_2020, reiner_modeling_2020}. One particular example that was reasonably effective in modeling behavioral changes during a pandemic was the closed-loop feedback in the compartmental SIR model presented by Perra et al.  \cite{perra_towards_2011}. The authors examined behavioral changes by modeling the rate at which individuals enter self-imposed quarantine dependent on the number of infectious individuals. 

Over the last year, the time history of new infected regional cases has fluctuated drastically and has posed significant challenges for the infectious disease modeling community.  A model that can represent the region/population specific COVID-19 cases accurately for the entire period of this pandemic has not yet been reported. We propose a mathematical model and a framework that incorporates the naturally occurring behavioral responses of a population to infectious cases coupled with possible additional behavioral changes exhibited during vaccination that has the ability to represent infection dynamics for the entirety of available COVID-19 case data in the United States (US) regions. We define ``level of caution" to represent a population's precautionary/safe behavior during an ongoing pandemic that results from a combination of increased social distancing, use of personal protection equipment, improved hygiene, and lockdown regulations. We also introduce ``sense of safety" to represent a population's return to normal, pre-pandemic behavior as more and more people are vaccinated. We introduce suitable mathematical forms to represent these two important behavioral aspects and incorporate these dynamic functions into our differential SIRDV model framework. 

Fitting our model to available daily new infection case data for four major US states (Massachusetts, California, Florida and South Dakota) and two major US cities (Atlanta and New York City) for the first year of the pandemic, we show that our modeling framework is versatile at capturing a large range of developments of the COVID-19 pandemic over time, and it provides valuable insights into each population's underlying social behavior. Introducing a vaccine to the population in our model, we analyzed the interaction between vaccine distribution rate and vaccine related additional behavioral responses of the population. We used our model to predict future trends for the pandemic with the advent of vaccine distribution. 


\section{Results}
We find that the time dependent infectious disease transmission rate $\beta(t)$ is best given by $ \beta=\beta_0 \,f_I \, f_V$, where $\beta_{0}$ is the population maximum infection transmission rate observed in the absence of any preventative societal measures. $f_I$ and $f_V$ are level of caution and sense of safety functions, respectively, proposed as: 
\begin{equation}
\begin{array}{l}
    f_I=e^{-d_I I} \\
    f_V=\dfrac{1}{f_I}+\left(1-\dfrac{1}{f_I}\right)e^{-d_V V} .\\
\end{array}
\label{f1_f2_eqns0}
\end{equation}
The function $f_I$ models caution in a population, where its individuals take measures to reduce disease transmission through social distancing, personal protective equipment, hygiene and local government mandates. The population's level of caution to the number of infectious cases is determined by a factor $d_I$, which was observed to change several times over a long duration in a given population due to changing population awareness and response, pandemic fatigue, seasons, and changing government mandates. These changes in sensitivity of the population to the number of infectious cases gives rise to the multiple peaks in the number of new infected cases observed nearly universally during the COVID-19 pandemic. $f_I$ approaches $0$ in the limiting case of very high values of level of caution factor $d_I$, reflecting extreme cautionary measures by the population against the pandemic and leads to negligible disease transmission. A $d_I$ value approaching $0$ gives $f_I$ as $1$, reflecting a population whose behavior is approaching pre-pandemic levels of minimal disease related precautionary actions. 

In addition, we included a competing sense of safety in our model, in which measures to reduce disease transmission are gradually decreased due to an increasing proportion of the population becoming vaccinated, offsetting the effects of a reduced transmission rate arising from an underlying level of caution. However, as modeled in equations \ref{f1_f2_eqns0}, the net transmission rate will never exceed the base maximum transmission rate $\beta_0$. As the sense of safety factor $d_V$ approaches a very small value (population not dropping its guards down due to vaccinations), the sense of safety function $f_V$ approaches 1 and $f_V$ has no effect on $\beta$. On the contrary, a high $d_V$ reflects an increased sense of safety, causing cautionary measures against the disease transmission to be significantly reduced, leading to $f_V = \frac{1}{f_I}$. In this case, infection-related level of caution is completely negated and the disease transmission rate $\beta$ approaches the population's highest transmission rate $\beta_0$.

\subsection{Infection data fit and interpretation for COVID-19 in the United States}

\begin{figure*}[tbhp]
    \centering
      \includegraphics[width=0.9\linewidth]{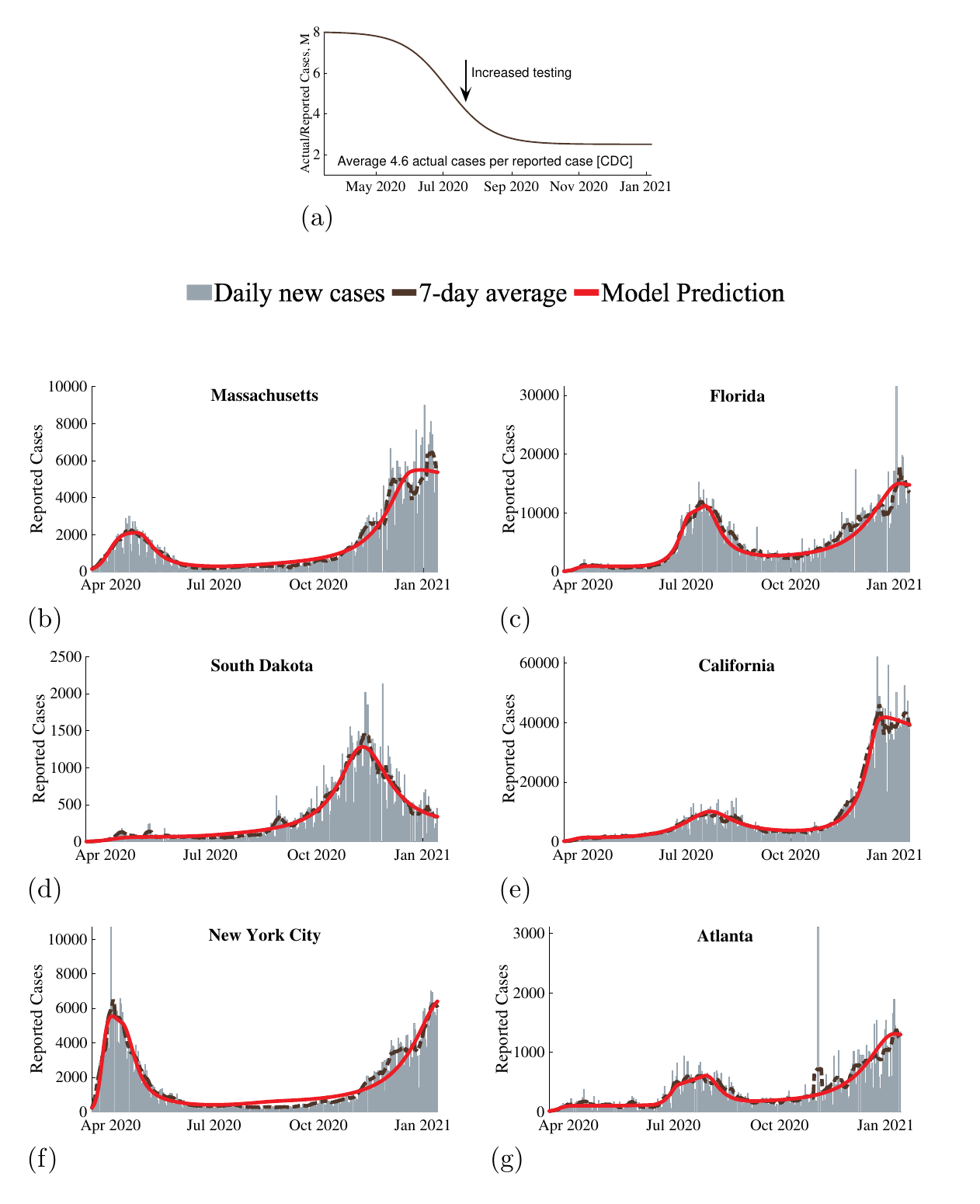} 
        \caption{Results of our behavioral model's fit to COVID-19 data across six populations in the United States. (a) Shows $M$ representing the ratio of actual infectious cases to reported cases.  Reported new infection case data versus model fit for (b) Massachusetts, (c) Florida, (d) South Dakota, (e) California, (f) New York City, and (g) Atlanta. Gray bars represent the daily reports of new COVID-19 cases. The dashed brown lines are the corresponding 7-day average. Our model's predictions for reported daily cases are shown as the solid red lines.}
    \label{pop_cases}
\end{figure*}

 We show that our modeling framework that mathematically incorporates the dynamic level of caution within a SIRDV differential framework (shown in Figure \ref{model_related}, represented by equations \ref{modelEquations} and discussed in detail in Section \ref{methods}) is able to fit and predict the entire COVID-19 case history\footnote{The same model works for both pre-vaccine and during vaccination periods. Before vaccines become available, the vaccinated population fraction $V$ in equation \ref{f1_f2_eqns0} stays zero giving $f_V=1$ and naturally leading to no effects from vaccine related sense of safety.} for the selected representative populations\footnote{For brevity we show results for selected key populations. The model can be applied to other regions/populations within or outside of the United States as well.} within the US. The model was fit to four US states (Massachusetts, California, Florida and South Dakota) and two major US cities (New York City and Atlanta). These regions were chosen to represent a variety of population densities and varying geographical locations. We accounted for the fact that the reported cases were lower than the actual infection cases in the population due to lack of testing and asymptomatic cases using a factor $M$. $M$ has a high value at the beginning of the pandemic due to lack of testing and reduces to a lower value as testing becomes more available. The simplified $M$ shown in Figure \ref{pop_cases}a is assumed following CDC's assessment that only 1 out of 4.6 COVID-19 cases were reported in the US for 2020. As shown in Figure \ref{pop_cases}, our behavioral model was able to accurately model and fit representative states and  cities across the United States with few parameters for each region. Estimated parameters for each region are shown in Table \ref{table:pop_params}. 

\begin{figure*}[htbp!]
    \centering
      \includegraphics[width=0.9\linewidth]{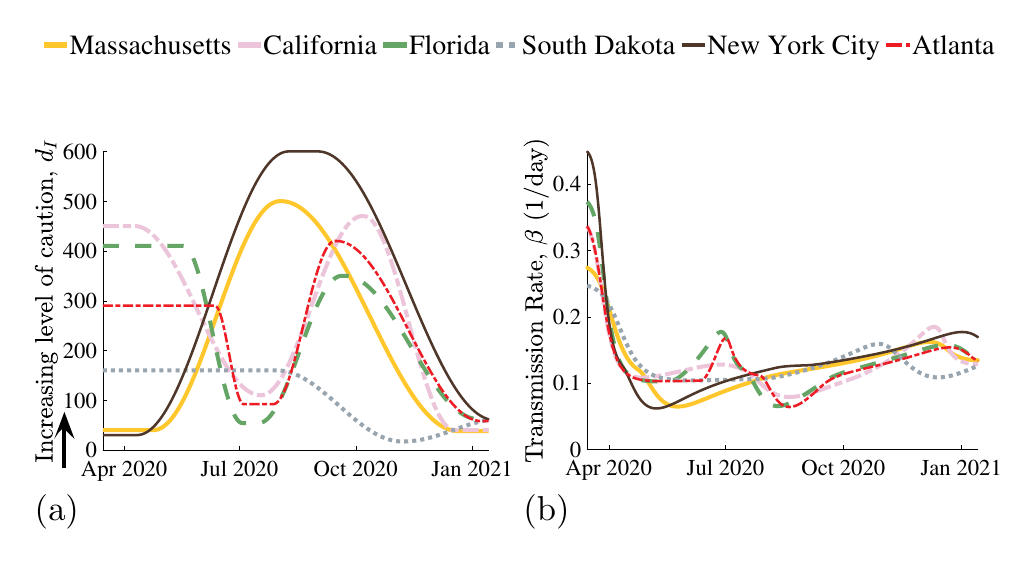} 
    \caption{(a) Level of caution factor $d_I$ and (b) disease transmission rate $\beta$ over the course of COVID-19 in six populations across the United States.  }
     \label{dI_beta}
\end{figure*}

The level of caution factor $d_I$ in equation \ref{f1_f2_eqns0} is shown in Figure \ref{dI_beta}a for different regions over the first year of the COVID-19 pandemic. A high level of caution factor indicates that the population was quick to adapt their behavior in response to an increase in infections by taking increasingly stringent measures to reduce their transmission rate. Level of caution in a specific population changes due to addition or removal of local government regulations, new information regarding the disease, seasonal changes in behavior, pandemic fatigue, news leading to additional fear or any other factor that causes widespread changes in behavior and disease transmission rate. As expected, the results show that sudden drops in level of caution factor $d_{I}$ tend to precede surges in new cases due to relaxed social measures. Conversely, a reduction in the influx of new cases will occur due to significant increase in $d_{I}$. This level of caution is independent of the baseline maximum transmission rate, $\beta_0$, and therefore provides a measure to compare social outlook towards the disease between different populations/regions. Time varying COVID-19 transmission rate $\beta$ for each of the selected regions is shown in Figure \ref{dI_beta}b. $\beta_{0}$ describes transmission in the earliest stages of the pandemic for a population, when knowledge of the disease and social measures against it were limited. Therefore, this value also describes the transmission that a specific population can be expected to return to when the precautions against the infection becomes minimal, either as infectious cases approaches zero or as social response to the disease becomes very low. In addition to the infectious disease inherent contagious characteristics, the  base transmission rate depends on factors such as population density, contact rate, and everyday pre-pandemic behavior of its individuals. Likely due to such factors, we found that bustling New York City, was on the higher end of the baseline transmission rate with basic reproductive ratio for New York City obtained to be $R_0 = \frac{\beta}{\gamma} = 4.5$), whereas a less densely populated state like South Dakota has a much lower baseline transmission rate and a lower $R_0 $ value of $2.5$. $R_0$ values for other regions can be found from Table \ref{table:pop_params}.

\begin{table}[h] \footnotesize
\begin{center}
\resizebox{\columnwidth}{!}{%
\begin{tabular}{|c| c |c| c| c |} 
 \hline
 Region & $\mu$ [\%] & 1/$\gamma$ [days] & $\beta_{0}$ [1/days] & $d_{I}$ \;\quad (Behavioral Period) \\ [0.5ex] 
 \hline\hline
 Massachusetts & & & 0.28 & \begin{tabular}{@{}l@{}}40 \,\;\quad (March 15 - Apr 24)\\500 \quad(Aug 2 - Aug 3) \\ 38 \,\;\quad (Dec 21 - )  \end{tabular}\\  
 \cline{1-1} \cline{4-5}
 California & \multirow{6}{*}{0.5} & \multirow{6}{*}{10} & 0.35 & \begin{tabular}{@{}l@{}}450 \quad (March 15 - Apr 9)\\110 \quad(Jul 18 - Jul 19) \\470 \quad(Oct 6 - Oct 7) \\ 40 \,\;\quad (Dec 18 - )  \end{tabular}\\
 \cline{1-1} \cline{4-5}
 South Dakota &   &  & 0.25 & \begin{tabular}{@{}l@{}} 160 \;\quad (March 15 - Jul 30)\\17 \;\;\quad(Nov 7 - Nov 8) \\ 63 \, \;\quad (Jan 26 - )  \end{tabular}\\ 
 \cline{1-1} \cline{4-5}
 Florida &  &  & 0.38 & \begin{tabular}{@{}l@{}} 410 \quad (March 15 - May 17)\\ 54 \,\;\quad(Jul 3 - Jul 16) \\ 350 \quad (Sep 20 - Sep 22) \\ 62 \,\;\quad (Jan 6 - )  \end{tabular}\\ 
 \cline{1-1} \cline{4-5}
 New York City &   &  & 0.45 & \begin{tabular}{@{}l@{}} 30 \,\;\quad (March 15 - Apr 11)\\600 \quad(Aug 9 - Sep 1) \\ 60 \,\;\quad (Jan 19 - )  \end{tabular}\\
 \cline{1-1} \cline{4-5}
 Atlanta &   &  & 0.35 & \begin{tabular}{@{}l@{}} 290 \;\quad (March 15 - Jun 11)\\92 \,\; \quad(Jul 4 - Jul 27) \\ 420 \;\quad (Sep 14 - Sep 15) \\ 58 \, \;\quad (Jan 8 - )  \end{tabular}\\
 \hline
\end{tabular}
}
\caption{Parameters for the four states and two cities for the 2020 COVID-19 pandemic. The initial basic reproduction number, $R_0$ = $\frac{\beta_0}{\gamma}$, represents the expected number of secondary infections if an infectious individual was placed in the population, without any social preventative measures. $d_I$ values represent its constant value in each of the corresponding behavioral periods. Smooth transitions between constant values of $d_I$ were applied.}
\label{table:pop_params}
\end{center}
\end{table}

To illustrate direct correlations between our model fit predictions and real life events, we take New York City as an example. Starting March 22, New York implemented the ``New York State on PAUSE” executive order, closing all non-essential businesses, canceling all non-essential gatherings, and mandated social distancing. This local government regulation is directly represented in the model results for level of caution which shows a significant increase in $d_I$ following this mandate (Figure \ref{dI_beta}a). Corresponding model results also show a sharp transition from one of the highest levels of disease transmission rates (Figure \ref{dI_beta}b) to one of the lowest levels of transmission rates following this government mandate. Between September 1, 2020 and January 15, 2021, the model based $d_I$ values show that the level of caution in New York City transitioned from one of the highest values to one of the lowest. When we examine real events, we find that starting September 2020, a series of citywide re-openings were introduced, including the opening of gyms, malls (at 50\% capacity), public K-12 schools, and indoor dining (25\% occupancy). This reopening coincided with the holiday season in the US at the end of the year 2020 and resulted in a significant spike in the new infection cases directly correlating with our model predictions. Therefore, the level of caution parameter $d_I$ is a metric that quantifies a population's behavior in response to an infectious disease outbreak; estimates of future $d_I$ values will allow predictions of new infectious cases.  

Overall, there are clear trends in COVID-19 cases captured by our model that directly relate to local government mandated health regulations. These results suggest that we may be able to incorporate possible behavioral changes into our model representing future government regulations, along with changes in vaccination rates, to predict infection outcomes.

\subsection{Future COVID-19 dynamics with vaccination}

\begin{figure*}[htbp!]
    \centering
      \includegraphics[width=0.9\linewidth]{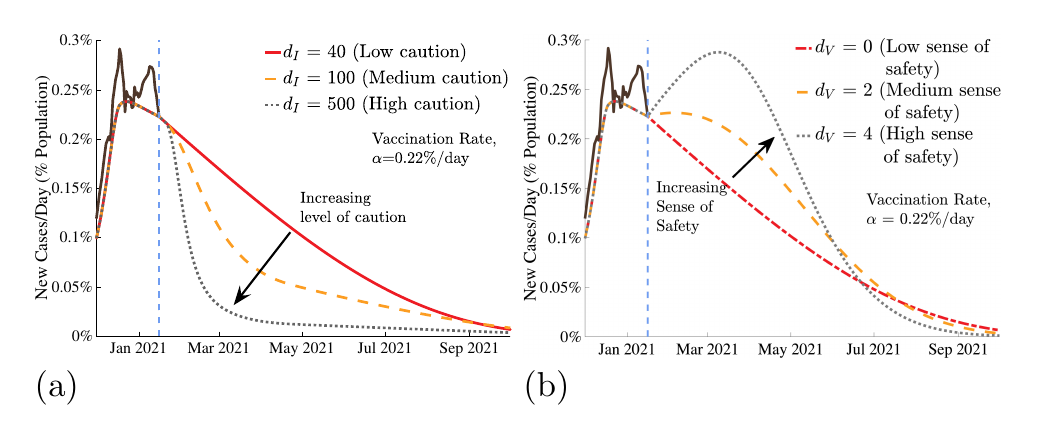} 
    \caption{Effect of (a) sense of caution factor, $d_I$, and (b) sense of safety factor, $d_V$ on pandemic trajectory. The dashed blue bar represents the introduction of a vaccine with 95\% effectiveness, at a fixed rate of 0.22\% of the population per day. Extreme values of $d_{I}$ represent the extremes of social behavior: the most responsive (black dotted) and the least (red solid).}
    \label{vary_dI_dV}
\end{figure*}

The model incorporates the effect of vaccination and the behavioral response of the population to growing number of people getting vaccinated due to the population's sense of safety. This sense of safety counteracts the underlying level of caution that a population always has in response to the number of infectious cases. Predictions from our model with the presence of vaccination show that the future trajectory of the pandemic will strongly depend on population’s behavior in response to the disease and vaccination. In Figure \ref{vary_dI_dV}, we show a range of potential infection outcomes for different levels of caution to the infection and different senses of safety due to vaccination. The selected range of $d_{I}$ and $d_V$ represent reasonable extremes of the level of caution and sense of safety factors. For the future trend predictions, we use starting values based on California's data, as a representative population (to avoid multiple curves and repetition) which has reasonable correlation with overall United States COVID-19 trends. The results, normalized as population fractions, provide critical COVID-19 future trends and insights that will be applicable to other regions as well. We have assumed a vaccine effectiveness $\eta$ of 95\% based on the initial estimates of the two leading vaccines \cite{Kim2021}. $\eta$ can be suitably selected for any other vaccination types or in the case of a different pandemic or endemic with different vaccine effectiveness. The results are shown for estimated actual cases. 

Figure \ref{vary_dI_dV}a models a population that does not alter their underlying behavior in response to introduction of a vaccine, and instead only responds to the increasing infectious cases by increasing personal safety measures. All simulated curves show a swift reduction in cases following the vaccine, though decisive action and stronger level of caution in response to the infection does show a considerable reduction in total infections. Figure \ref{vary_dI_dV}b shows a scenario in which the population responds to the introduction of a vaccine by relaxing the social measures meant to slow the transmission. Although this sense of safety and increased normalcy may be a natural response to vaccines becoming more available, our predictions show that unfailing and continuing commitment to social preventative measures can significantly reduce the total number of future infections and even prevent a new surge and new peak that can happen if the population relaxes too soon. Note, regardless of the value of the sense of safety factor $d_{V}$, the number of new cases per day drops to zero around the same time; however, the peak number of cases while progressing toward this point is very different for different $d_{V}$ values. 

We treat vaccination as a one-time event where an individual would receive an entire dose of the vaccine, despite the fact that some current vaccines require two doses that must be delivered at different times \cite{news_biden_nodate}\footnote{Given the much longer time scale of COVID-19 predictive curves, compared to the time gap between the two doses of m-RNA vaccines, it is reasonable to model vaccination rate by ignoring the time gap between the two doses and take the two doses combined as a single completed vaccine without significant loss of predictive accuracy.}. Unless otherwise noted, all vaccine distribution in this paper was modeled at a fixed rate of 0.22\% of the population per day (0.44\% per day in terms of doses for two dose vaccines). This number represents approximately 1.5 million vaccine doses that are currently administered in the US.  Because of the asymptomatic and unreported cases, the constant rate vaccination was applied proportionally to both the susceptible ($\alpha_s$) and recovered ($\alpha_R$) populations. For the remainder of the simulations, we  assume that sense of caution, $d_{I}$, remains fixed at one of it's recent (January) estimate ($d_{I}=40$).

\begin{figure*}[htbp!]
    \centering
      \includegraphics[width=0.95\linewidth]{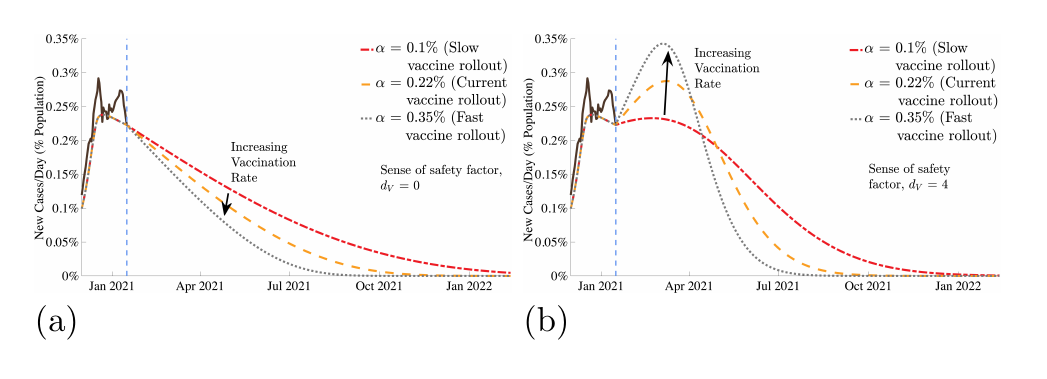} 
      \caption{Impact of vaccination rate, $\alpha$, with two social responses to the vaccine. The population in (a) shows no changes in behavior in response to the vaccine, while the population in (b) relaxes its preventative measures as the vaccine is distributed.}
      \label{alphavals}
\end{figure*}

The effects of the vaccination rate $\alpha$ were examined (Figure \ref{alphavals}). We chose three values of $\alpha$: the current rate of vaccination of $0.22\%$ of the population per day \cite{mass_vaccinations}, a low ($0.10\%$ per day), and a high rate ($0.35\%$ per day). Note, the vaccination rate of $0.1\%$ is not expected in US but is shown to illustrate the consequences of low vaccination rate. 
As the vaccine distribution rate increases, the number of cases per day tends to zero quickly. However, as is shown in Figure \ref{alphavals}a and \ref{alphavals}b, the population's social response to the vaccine, $d_V$, has a significant effect on the pandemic trajectory. In cases where preventative measures were abandoned more quickly and the population had an increased sense of safety in response to the vaccines (high values of $d_V$), increased vaccination rates still result in cases quickly tending toward zero, but before this happens, the number of cases per day increases rapidly. This behavior worsens as $d_{V}$ increases, and this sharp increase occurs earlier as vaccination rate $\alpha$ increases. Our results show that in the US, COVID-19 can be reasonably controlled by late summer of 2021 proceeding with the currently planned vaccination rate. 

Our results elucidate the importance of local health and government authorities becoming aware of the fact that the sense of safety and vaccine distribution rate are related parameters. As has been shown, a faster vaccination rate significantly decreases the duration of the pandemic. However, if authorities intend to distribute a vaccine very quickly, they must be extra cognizant of the population's behavioral response to it, as population relaxing it's cautious practices could result in a noticeable increase in cases post vaccine rollout. If neglected, this peak under extreme circumstances could be disastrous. Therefore, based on our results, we recommend that proper disease transmission mitigating behavior be maintained, while welcoming a fast vaccine distribution rate. 

\begin{figure}[h]
    \centering
    \includegraphics[width=0.46\textwidth]{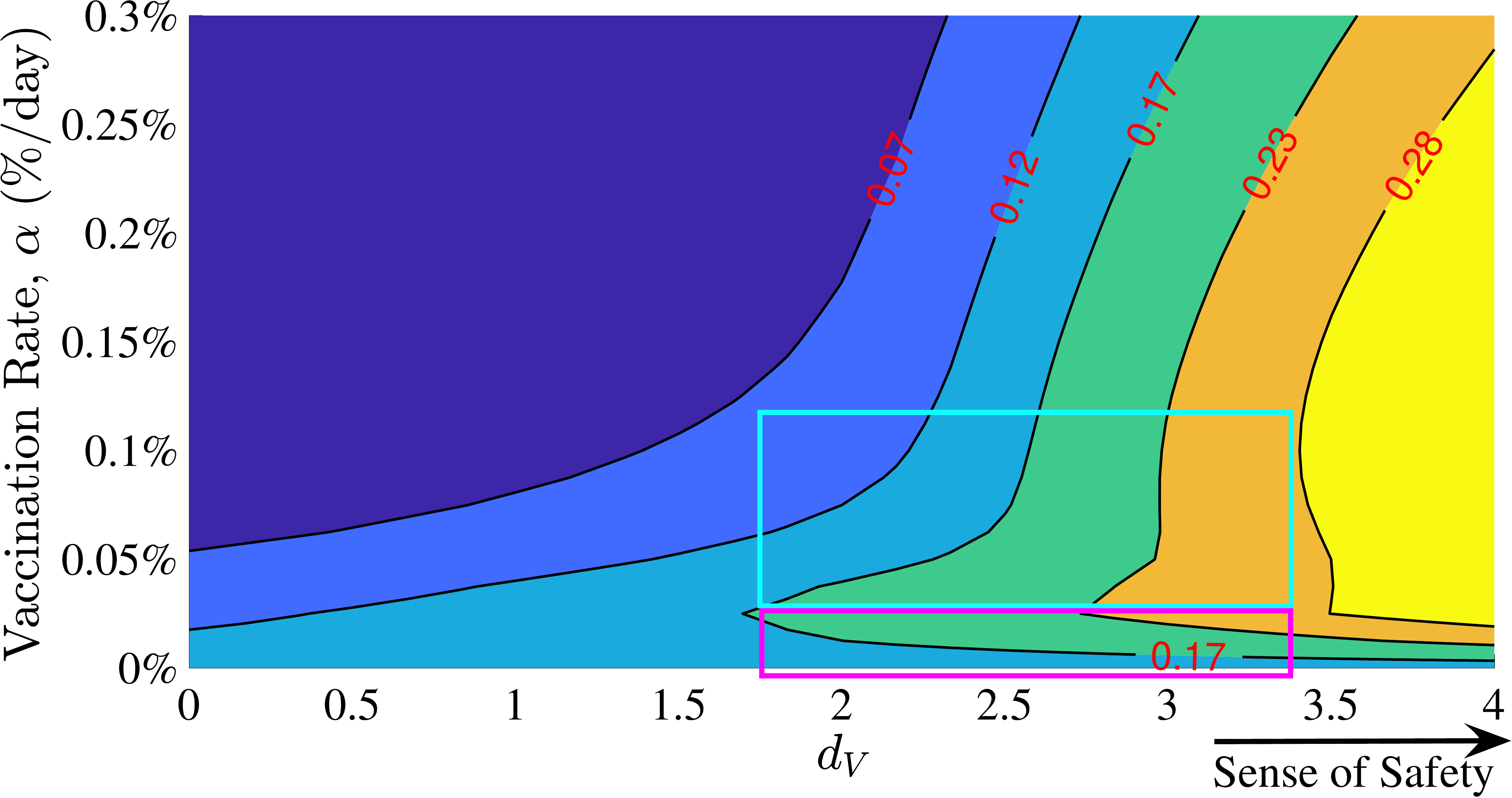}
    \caption{Contour plot of the sense of normalcy, $d_{V}$, and vaccination rate, $\alpha$, versus the total infections since the start of vaccination (red values shown as proportion of the total population). Shown for a high level of caution case ($d_{I}$=1000). Pink box shows a possibility of increasing number of total infected cases with increasing vaccination rate at very high sense of safety, and the teal box shows that once the vaccination rate crosses a threshold, the total number of infection cases drop with vaccination rate even at a very high sense of safety.}
    \label{fig:tot_inf_contours}
\end{figure}

To further quantify the relative effects of the sense of safety and the vaccine distribution rate on the total number of infectious cases,  the total number of individuals infected (as population fraction) after the start of vaccination were plotted with respect to $\alpha$ and $d_{V}$ in Figure \ref{fig:tot_inf_contours}. This was done for a special case of a very high level of caution during vaccine rollout. As expected, for a given value of vaccination rate, the number of total infected cases increases as the sense of safety factor increases along the x-axis. This behavior is especially pronounced for low vaccination rates, when large increases in the sense of safety can result in significant numbers of total infections, up to 28$\%$ as is shown in the yellow region of Figure \ref{fig:tot_inf_contours}). 
Also, note that for very high value of $d_V$, as the vaccination rate increases from 0, the number of infected cases quickly increases and then start to decrease again (pink box). For a very slow vaccination rate, vaccinated population dependent behavioral effects are limited due to our proposed relation for the sense of safety function $f_{V}$, but quickly increase as the vaccinated individuals increase. This explains the increase in the total infections as one travels vertically in the pink box in Figure \ref{fig:tot_inf_contours}. However, total infections then begin to decrease due to a critical vaccination rate being achieved, shown in the teal box. This reinforces the argument for the necessity of maximizing vaccine distribution; low vaccination rates can lead to behavior-related spikes in total cases, but these effects are mitigated as widespread vaccination outweighs these behavioral factors. 

\begin{figure}[h]
    \centering
    \includegraphics[width=0.46\textwidth]{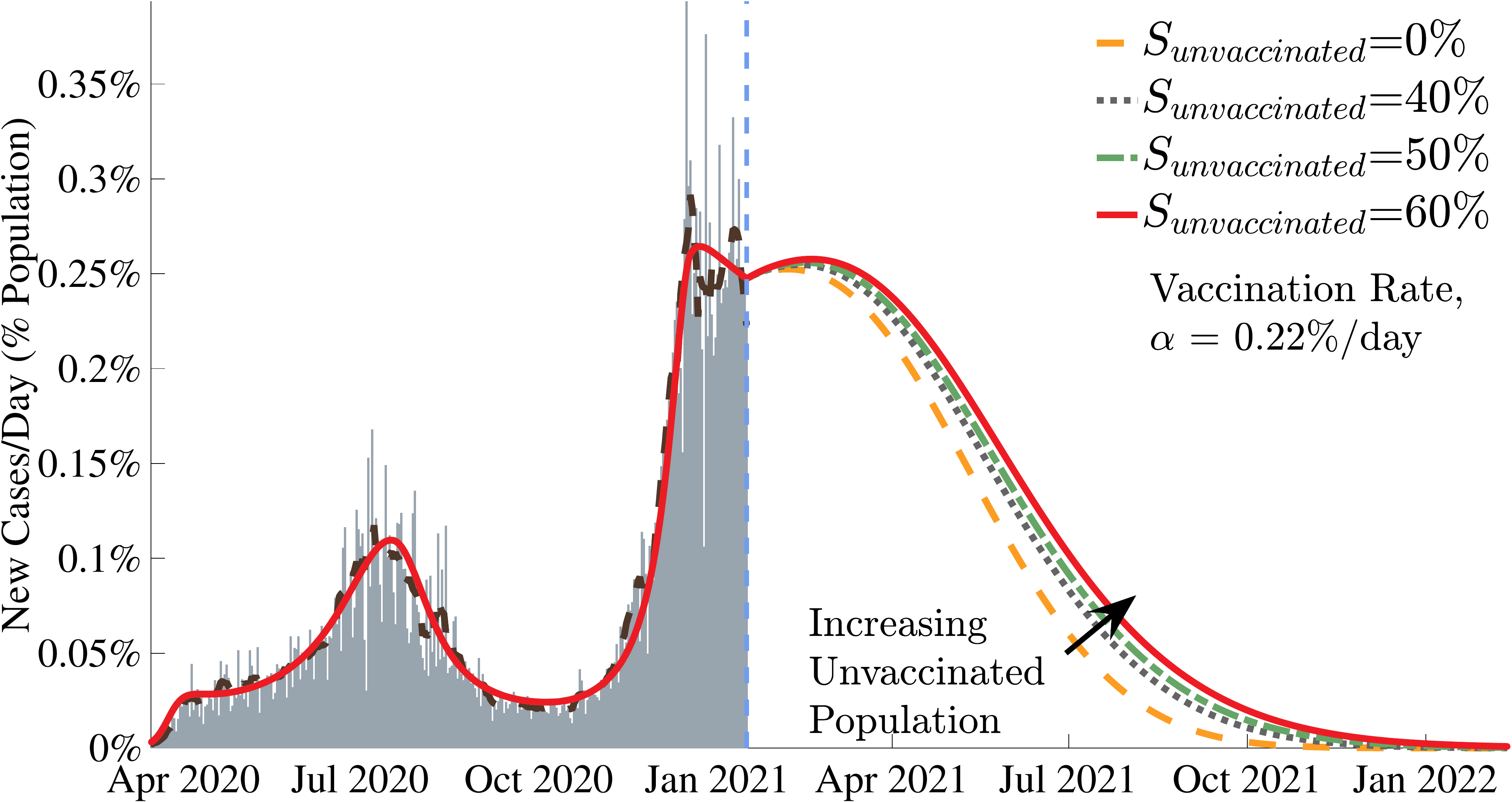}
    \caption{Results for certain fractions of the population remaining unvaccinated. The entire COVID-19 case history and future predictions are shown in terms of estimated actual new infection cases for California as a representative example. Gray bars represent the daily reports of new COVID-19 cases. The dashed brown lines are the corresponding 7-day average. Our model predictions for the entire duration are also shown as red, green, grey and yellow curves.}
    \label{fig:vacc_stop_d2}
\end{figure}

We can expect some portion of the population to be unwilling or unable to receive a COVID-19 vaccine. The effect of the size of this group on the duration and severity of the pandemic was examined in Figure \ref{fig:vacc_stop_d2}. If large populations refuse vaccination, the duration of the pandemic can be increased by several months. 

\section{Method}
\label{methods}
\subsection{Modeling approach}

\begin{figure*}[htbp!]
    \centering
      \includegraphics[width=0.8\linewidth]{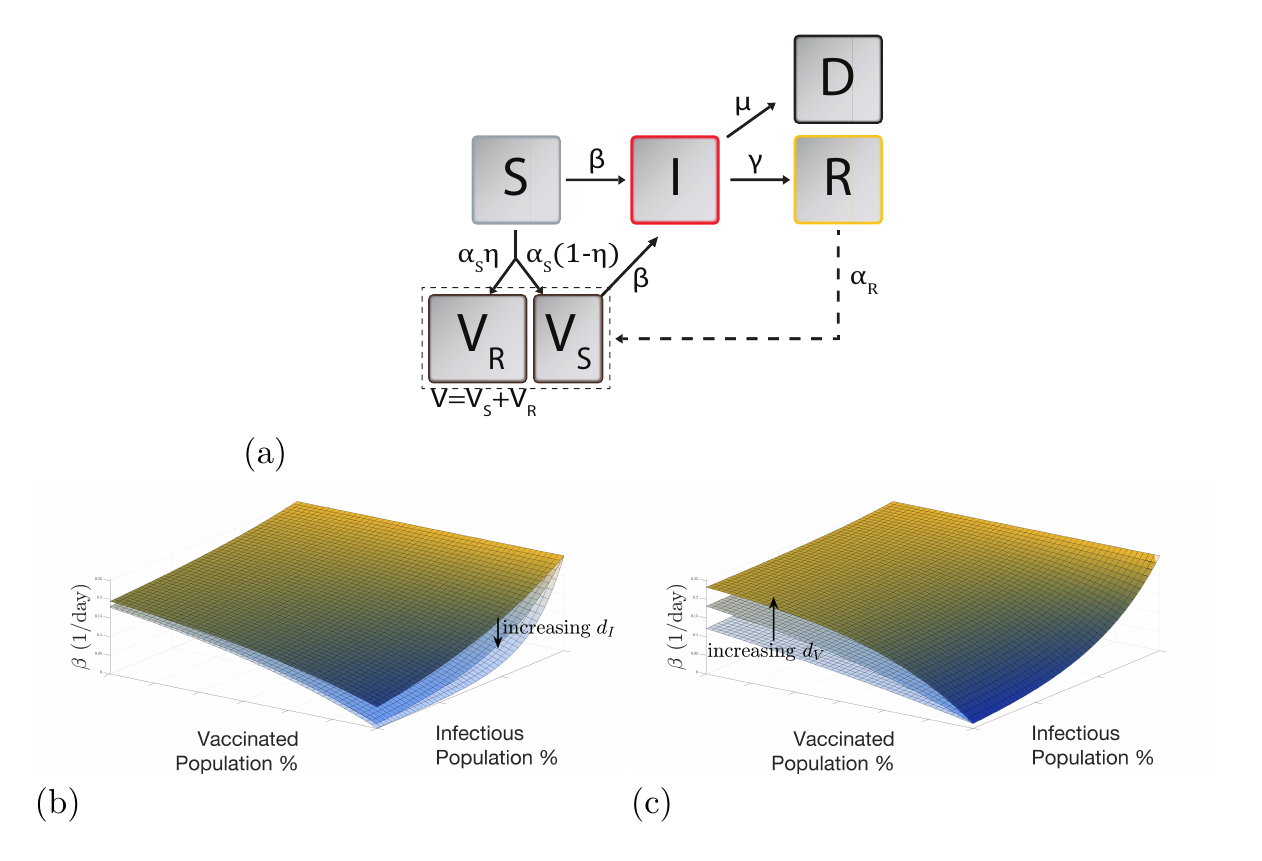} 
    \caption{(a) SIRDV compartment model with susceptible ($S$), vaccinated ($V$), infectious ($I$), recovered ($R$), and deceased ($D$) populations during an infection outbreak. Flow between between compartments is shown with arrows. (b) and (c) Show variations in disease transmission rate $\beta$, due to social response to infectious and vaccinated populations. (b) $\beta$ for a range of level of caution factor $d_I$ (100, 200, 400), and (c) $\beta$  for a range of sense of safety factor $d_V$ (1, 2, 4).} 
    \label{model_related}
\end{figure*}

As shown in Figure \ref{model_related}a, our model extends the general SIRD framework by adding the effect of vaccination and incorporating behavior based dynamics as an important capability specific to our study. The model consists of five compartments: Susceptible (S), Vaccinated (V), Infectious (I), Recovered (R), and Deceased (D). Here S,V, I, R, and D represent time dependent fractional variables with respect to the total population of the region of interest. Beginning in the susceptible compartment, individuals can follow the standard infection pathway through the infectious compartment then to either recovered or deceased. Alternatively, they can enter the vaccinated compartment following a fixed rate of vaccination $\alpha$, where depending on the vaccine efficacy $\eta$, a subset of the vaccinated population $V_s$ can become infected again. The remainder of the vaccinated group $V_R$ is successfully vaccinated and have no risk of becoming infected. The known reinfection rate is very low and it's effect can be neglected for the timescale of our study. Note that given the uncertainty in how rapidly the vaccines will be deployed in the future, for our predictions, we have used constant vaccination rates and have shown sensitivities to different vaccination rates. Time dependent vaccination rates can be easily implemented by selecting a suitable function for $\alpha(t)$ in our model. Our SIRDV model for a region/population is described by the following equations:
\begin{equation}
\begin{array}{l} \vspace{0.03in}
    \dfrac{dS}{dt}=-\beta SI-\alpha \\ \vspace{0.03in}
    \dfrac{dV}{dt}=-\beta\left(1-\eta\right)V I+\alpha \\\vspace{0.03in}
    \dfrac{dR}{dt}=\gamma I \\\vspace{0.03in}
    \dfrac{dI}{dt}=\Big(\beta S+\beta\left(1-\eta\right)V-\mu\gamma-\gamma\Big)I \\\vspace{0.03in}
    \dfrac{dD}{dt}=\mu\gamma I\\\vspace{0.03in}
\end{array}
\label{modelEquations}
\end{equation}
where $\beta$ represents the dynamic transmission rate, $\mu$ represents the mortality rate, and $\gamma$ represents the recovery rate. The family of curves represented by this set of equations with constant parameters is considerably limited as it assumes that the population does not change its behavior at all over the course of the outbreak. 

The significant differences and variations in disease transmission across different populations and over the course of the COVID-19 pandemic have shown that an understanding and modeling of dynamic population behavior changes is critical in predicting a real-world pandemic. To model these population behavioral attributes, we have incorporated a simple framework for a behavior-based, time-dependent net disease transmission rate $\beta$ that is dependent on both the current infectious and vaccinated populations. With $\beta_{0}$ as population maximum transmission rate, $f_I (\le 1)$ level of caution function and $f_V (1 \le f_V \le \frac{1}{f_I})$ as sense of safety function, we propose the following mathematical forms for behavior dependent transmission rate:

\begin{equation}
\begin{array}{l}
    f_I=e^{-d_I I} \;,\\
    f_V=\dfrac{1}{f_I}+\left(1-\dfrac{1}{f_I}\right)e^{-d_V V} \;,\vspace{0.04in}\\ 
    \beta=\beta_0 \,f_I \, f_V \;.
\end{array}
\label{f1_f2_eqns}
\end{equation}

The resultant effects of infectious and vaccinated populations on disease transmission rate $\beta$ are shown in Figures \ref{model_related}b and \ref{model_related}c for a range of $d_I$ and $d_V$ values. As shown, transmission rate decays to a smaller value at high infectious populations due to more cautionary and preventive actions with a higher level of caution. This decay slows significantly as a higher percentage of the population gets vaccinated due to the sense of safety from vaccination. The sensitivity of $\beta$ to infectious population size is determined by the population's $d_I$, while the extent to which preventative measures are abandoned due to vaccine distribution is determined by $d_V$. Note, from the mathematical form in equation \ref{f1_f2_eqns}, in the absence of vaccines, $V=0 \implies f_V = 1$, which is physically and intuitively correct. All the model parameters are described in Table \ref{table:params}. 

\begin{table}[htbp!]\footnotesize
\begin{center}
\resizebox{\columnwidth}{!}{%
\begin{tabular}{|c l|} 
 \hline
 Parameter & Description\\ [0.5ex] 
 \hline\hline
 $\mu$ & mortality due to COVID-19\\
 \hline
  $\gamma$ & recovery rate—inverse of the average infectious period\\
 \hline
  $\beta_{0}$ & \begin{tabular}{@{}l@{}}baseline transmission rate—product of the transmission \\probability and contact rate in a relatively unrestricted \\population before any behavioral response to the disease occurs \end{tabular}\\
 \hline
 $d_{I}$ & \begin{tabular}{@{}l@{}}sense of caution factor—at higher values, the population will \\tend to decrease their contact rate and take more preventive \\measures more quickly in response to a high infectious population\end{tabular}\\
 \hline
 $\alpha$ & vaccination rate\\
 \hline
$\eta$ & vaccine effectiveness\\
 \hline
 $d_{V}$ & \begin{tabular}{@{}l@{}}sense of safety factor—at higher values, the population will \\tend to reduce personal safety measures more quickly in response to a\\ high vaccinated population. This effect counteracts the effects of $d_{I}$\end{tabular}\\
 \hline
 M & \begin{tabular}{@{}l@{}} a factor representing ratio of actual cases to  reported cases \\due to asymptomatic cases and limited COVID-19 testing availability\end{tabular}\\
 \hline
\end{tabular}
}
\caption{Parameters in the model.}
\label{table:params}
\end{center}
\end{table}

\subsection{Model fit to specific regions}

Combining equations \ref{modelEquations} with our dynamic behavior model in equations \ref{f1_f2_eqns}, we are able to fit the complex, multimodal infection curves observed during the course of the COVID-19 pandemic.

We take the disease mortality rate and infectious period, $\mu$ and $\gamma$ (inverse of $\gamma$ represents infectious period) as constants. The baseline transmission rate $\beta_{0}$ was determined to be the parameter which was able to best fit the first rise in cases, where there was limited social response to the rise in infections. To represent the behavioral changes that were evident in the multiple peaks of the pandemic, we introduced multiple behavioral regions for each population. The model fit shows that each region had different level of caution factors (infection responsiveness) $d_{I}$, which provides an estimate of how public perception of the disease varied in each region over the course of the pandemic. Each behavioral response is represented by a fixed $d_I$ and smooth transitions were implemented via cosine interpolation, as displayed in Figure \ref{dI_beta}a. To reduce the risk of overfitting our model, we limited the number of behavioral response changes for each location and found that, for the locations that were selected, a minimum of either three or four behavioral regions was sufficient to accurately represent the complete reported infection data. The differential equation model was implemented in MATLAB. 

Model parameters were fit to daily new cases time series of four states and two cities: Massachusetts, California, Florida, South Dakota, New York City, and Atlanta. These regions were chosen to represent a variety of population densities, locations across the US, and responses to the pandemic. To determine the parameters, we used MATLAB's bound-constrained optimization function \cite{fminsearch}, minimizing the root mean square error between our model predictions and the reported number of cases per day. Simulations for each population began on March 15 and had an initial infected population equal to the number of new cases in the previous $\frac{1}{\gamma}$ days, to estimate those who were still in the infectious period.

The reported infection case data is limited by the fact that tests were not universally available during the beginning stages of the pandemic. Additionally, certain infected populations did not show any symptoms \cite{nogrady_what_2020}, but still may have transmitted the disease, further complicating estimates of new cases. To compare the actual case predictions from the model against the reported case, we introduced a factor M, which represents the number of actual cases per reported case. At the early stages of the pandemic, the awareness and testing was lacking but later on it improved significantly. We account for this by using a value of $M = M_0 $ for the initial stages of the pandemic which transition to a lower value of $M= M_f $ well into the pandemic when testing becomes widely available. The transition between the two values of $M$ is taken to be smooth using a sigmoidal function and is shown in Figure \ref{pop_cases}a.  The variation of $M$ with time can be represented with

\begin{equation}
    M=M_f+\frac{M_0-M_f}{1+e^{\delta_M(t - t_{M})}},
\label{M_eqn}
\end{equation}
where  $\delta_M$ and $t_{M}$ describe the smooth transition from $M_0$ to $M_f$. 

\subsection{Disease dynamics with vaccination}
After fitting the model to reported real-world data, the effects of vaccination and its potential effects on behavioral response (sense of safety) were examined.  Specifically, we chose the state of California as a representative case, which displayed an average infectivity rate of COVID-19 among the states and cities that we surveyed and an infection curve that was somewhat representative of the overall United States. Varying the vaccine distribution rate $\alpha$, the sense of safety parameter $d_{V}$, and the fraction of unvaccinated individuals $S_{unvaccinated}$ in our model, we evaluated the progression of the disease by examining the predicted number of cases reported per day in the future.  The predicted results are presented in Figure \ref{vary_dI_dV} through \ref{fig:vacc_stop_d2}.

\section{Closing Remarks}

We have developed an infectious disease dynamics model which accounts for behavioral changes in a population considering level of caution due to growing infectious individuals as well as a counteracting trend towards increasing normalcy, relaxing precautionary measures due to a sense of safety from increasing vaccine deployment. Our mathematical model  accurately captures the infection trends for the first year of the COVID-19 pandemic for all of the US regions examined with a small number of parameters. A comparison of model parameters between different regions allows comparative insights between them. It demonstrates direct relationships between population behavior model parameters and major government actions that impact population behavior. It allows measurement of several important population and infectious disease specific quantities including highest disease transmission rate $\beta_0$, disease transmission rate at any given time $\beta$, and a measure of population's behavior to reduce the disease transmission through parameter $d_I$, where, in the absence of significant vaccination, $\frac{d_I} {100} >> 1$ indicates safe response, and $\frac{d_I} {100}< 1$ represents lack of caution. 

We found that although faster vaccine rollout will bring the COVID-19 end more quickly, there exist scenarios where fast vaccine rollout can give false sense of safety to the population, which will lead to a large short-term increase in infectious cases. Prudence is required on the part of authorities to understand, predict, and limit any potential surge by increasing encouragement of all cautionary measures to prevent the spread of the virus. Our results indicate that in the United States, current COVID-19 strain can be reasonably controlled by August 2021 with the planned fast vaccination roll out. 

While our model is built on significant physical insights, population's future behavioral aspects, presence of unknown asymptomatic cases, a lack of exact knowledge about future vaccination rates, and other factors create some uncertainties. Therefore, although the quantitative predictions from our study are important, all possible uncertainties should be considered.  The results allow new insights into future COVID-19 trends and sensitivity of pandemic dynamics to various behavioral and other model parameters. As more exact information becomes available, new data can be directly incorporated in our model to produce more accurate results. Our model provides a new framework for predicting infection dynamics of future epidemics and pandemics.


\phantomsection
\subsection*{Acknowledgments} 

\addcontentsline{toc}{section}{Acknowledgments} 

Authors would like to thank John Antolik  and Thomas Bohac for discussions during the early phase of this research project.

\subsection*{Funding Information}
No external funding was used for this study. 

\subsection*{Author Contributions}
T.U. and Z.L. performed simulations and helped develop the model. V.S. conceptualized and led this research study and the development of the model. 

\subsection*{Competing Interest}
The authors declare no competing interests.

\subsection*{Data Archival}
COVID data was obtained from the Center for Systems Science and Engineering (CSSE) COVID-19 Data Repository at Johns Hopkins University \cite{dong_interactive_2020}. Estimates for the total population of each region were obtained from the United States Census Bureau \cite{bureau_county_nodate}.

\subsection*{Materials and Correspondence}
All correspondence should be addressed to Vikas Srivastava (vikas\_srivastava@brown.edu)


\footnotesize
\phantomsection
\bibliographystyle{unsrt}
\bibliography{BibliographyCovid} 


\end{document}